\documentstyle[11pt]{article}

\begin{document}

\title{A topological study of contextuality and modality in quantum mechanics}

\author{{\sc Graciela
Domenech}\thanks{%
Fellow of the Consejo Nacional de Investigaciones Cient\'{\i}ficas
y T\'ecnicas (CONICET)} $^{1, 3}$,\  \  {\sc Hector Freytes}
$^{2}$\ {\sc and} \ {\sc Christian de Ronde} $^{3, 4}$}

\maketitle

\begin{center}

\begin{small}
1. Instituto de Astronom\'{\i}a y F\'{\i}sica del Espacio (IAFE)\\
Casilla de Correo 67, Sucursal 28, 1428 Buenos Aires, Argentina\\
2. Dipartimento di Scienze e Pedagogiche e Filosofiche -
Universita degli Studi di Cagliari - Via Is Mirrionis 1, 09123,
Cagliari - Italia\\ 3. Center Leo Apostel (CLEA)\\ 4. Foundations
of the Exact Sciences (FUND) \\ Brussels Free University -
Krijgskundestraat 33, 1160 Brussels - Belgium
\end{small}
\end{center}

\begin{abstract}

\noindent

Kochen-Specker theorem rules out the non-contextual assignment of
values to physical magnitudes. Here we enrich the usual
orthomodular structure of quantum mechanical propositions with
modal operators. This enlargement allows to refer consistently to
actual and possible properties of the system. By means of a
topological argument, more precisely  in terms of the existence of
sections of sheaves, we give an extended version of Kochen-Specker
theorem over this new structure. This allows us to prove that
contextuality remains a central feature even in the enriched
propositional system.

\end{abstract}

\begin{small}


\end{small}

\bibliography{pom}

\newtheorem{theo}{Theorem}[section]

\newtheorem{definition}[theo]{Definition}

\newtheorem{lem}[theo]{Lemma}

\newtheorem{prop}[theo]{Proposition}

\newtheorem{coro}[theo]{Corollary}

\newtheorem{exam}[theo]{Example}

\newtheorem{rema}[theo]{Remark}{\hspace*{4mm}}

\newtheorem{example}[theo]{Example}

\newcommand{\proof}{\noindent {\em Proof:\/}{\hspace*{4mm}}}

\newcommand{\qed}{\hfill$\Box$}

\newcommand{\ninv}{\mathord{\sim}} 

\section{Introduction}

Modal interpretations of quantum mechanics (Dieks, 1988; Dieks,
1989; Dieks, 2005; van Fraassen, 1991) face the problem of finding
an objective reading of the accepted mathematical formalism of the
theory, a reading ``in terms of properties possessed by physical
systems, independently of consciousness and measurements (in the
sense of human interventions)'' (Dieks, 2005). These
interpretations intend to consistently include the possible
properties of the system in the discourse looking for a new link
between the state of the system and the probabilistic character of
its properties, thus sustaining that the interpretation of the
quantum state must contain a modal aspect. The name {\it modal
interpretation} was for the first time used by B. van Fraassen
(van Fraassen, 1981) following {\it modal logic}, precisely the
logic that deals with possibility and necessity. Within this
frame,  a physical property of a system means ``a definite value
of a physical quantity belonging to this system; i.e., a feature
of physical reality'' (Dieks, 2005). As usual, definite values of
physical magnitudes correspond to yes/no propositions represented
by orthogonal projection operators acting on vectors belonging to
the Hilbert space of the (pure) states of the system (Jauch,
1996).

Formal studies of modal interpretations of quantum logic exist
which are similar to the modal interpretation of the
intuitionistic logic (Dalla Chiara, 1981; Goldblatt, 1984). A
description of that those kind of approaches may be found in
(Dalla Chiara {\it et al.}, 2004). At first sight, it may be
thought that the enrichment of the set of (actual) propositions
with modal ones could allow to circumvent the contextual character
of quantum mechanics. We have faced the study of this issue and
given a Kochen-Specker type theorem for the enriched lattice
(Domenech {\it et al.}, 2006). In this paper, we give a
topological version of that theorem, i.e. we study contextuality
in terms of sheaves.

\section{Basic Notions}

We recall from (Goldblatt, 1986; Mac Lane and Moerdijk, 1992) and
(Maeda and Maeda, 1970) some notions of sheaves and lattice theory
that will play an important role in what follows. First, let $(A,
\leq)$ be a poset and $X\subseteq A$.  $X$ is a {\it decreasing
set} iff for all $x\in X$, if $a\leq x$ then $a\in X$. For each
$a\in A$ we define the {\it principal decreasing set} associated
to $a$ as $(a] = \{x\in A: x\leq a \}$. The set of all decreasing
sets in $A$ is denoted by $A^+$, and it is well known that $(A^+,
\subseteq)$ is a complete lattice, thus $\langle A, A^+  \rangle $
is a topological space. We observe that if $G\in A^+$ and $a\in G$
then $(a]\subseteq G$. Therefore $B = \{(a]: a\in A \}$ is a base
of the topology $A^+$ which we will refer to as the {\it canonical
base}.  Let $I$ be a topological space. A {\it sheaf} over $I$ is
a pair $(A, p)$ where $A$ is a topological space and $p:A
\rightarrow I$ is a local homeomorphism. This means that each
$a\in A$ has an open set $G_a$ in $A$ that is mapped
homeomorphically by $p$ onto $p(G_a) = \{p(x): x\in G_a\}$, and
the latter is open in $I$. It is clear that $p$ is a continuous
and open map. {\it Local sections} of the sheaf $p$ are continuous
maps $\nu: U \rightarrow I$ defined over open proper subsets $U$
of $I$ such that $p \nu = 1_U$. In particular we use the term {\it
global section} only when $U=I$.

In a Boolean algebra $A$, congruences are identifiable to certain
subsets called {\it filters}. $F\subseteq A$ is a filter iff it
satisfies: if $a\in F$ and $a\leq x$ then $x\in F$ and if $a,b \in
F$ then $a\land b \in F$. $F$ is a proper filter iff $F\not=A$ or,
equivalently $0\not \in F$. If $X\subseteq A$, the filter $F_X$
generated by $X$ is the minimum filter containing $X$. A proper
filter $F$ is {\it maximal} iff the quotient algebra $A/F$ is
isomorphic to ${\bf 2}$, being ${\bf 2}$ the two elements Boolean
algebra. It is well known that each proper filter can be extended
to a maximal one.

We denote by ${\cal OML}$ the variety of orthomodular lattices.
Let $L=\langle L,\lor,\land, \neg, 0, 1\rangle$ be an orthomodular
lattice. Given $a, b, c$ in $L$, we write: $(a,b,c)D$\ \   iff
$(a\lor b)\land c = (a\land c)\lor (b\land c)$; $(a,b,c)D^{*}$ iff
$(a\land b)\lor c = (a\lor c)\land (b\lor c)$ and $(a,b,c)T$\ \
iff $(a,b,c)D$, (a,b,c)$D^{*}$ hold for all permutations of $a, b,
c$. An element $z$ of a lattice $L$ is called  {\it central} iff
for all elements $a,b\in L$ we have\ $(a,b,z)T$. We denote by
$Z(L)$ the set of all central elements of $L$ and it is called the
{\it center} of $L$. $Z(L)$  is a Boolean sublattice of $L$ (Maeda
and Maeda, 1970; Theorem 4.15).

\section{Sheaf-theoretic view of contextuality}

Let ${\mathcal H}$ be the Hilbert space associated to the physical
system and $L({\mathcal H})$ be the set of closed subspaces on
${\mathcal H}$. If we consider the set of these subspaces ordered
by inclusion, then $L({\mathcal H})$ is a complete orthomodular
lattice (Maeda and Maeda, 1970). It is well known that each
self-adjoint operator $\bf A$ that represents a physical magnitude
$A$ may be associated with a Boolean sublattice $W_A$ of
$L({\mathcal H})$. More precisely, $W_A$ is the Boolean algebra of
projectors ${\bf P}_i$ of the spectral decomposition ${\bf
A}=\sum_{i} a_i {\bf P}_i$. We will refer to $W_A$ as the spectral
algebra of the operator $\bf A$. Any proposition about the system
is represented by an element of $L({\mathcal H})$ which is the
algebra of quantum logic introduced by G. Birkhoff and J. von
Neumann (Birkhoff and von Neumann, 1936).

Assigning values to a physical quantity {\it A} is equivalent to
establishing a Boolean homomorphism $v: W_A \rightarrow {\bf 2}$
(Isham, 1998). Thus, it is natural to consider the following
definition which provides us with a {\it compatibility condition}:

\begin{definition}\label{ccon}
{\rm
Let $(W_i)_{i\in I}$ be the family of Boolean sublattices of
$L({\mathcal H})$. A global valuation over $L({\mathcal H})$ is a
family of Boolean homomorphisms $(v_i: W_i \rightarrow {\bf
2})_{i\in I}$ such that $v_i\mid W_i \cap W_j = v_j\mid W_i \cap
W_j$ for each $i,j \in I$.}

\end{definition}

Kochen-Specker theorem (KS) precludes the possibility of assigning
definite properties to the physical system in a non-contextual
fashion (Kochen and Specker, 1967). An algebraic version of KS
theorem is given by (Domenech and Freytes, 2005; Theorem 3.2):

\begin{theo}\label{CS2}
If $\mathcal{H}$ be a Hilbert space such that $dim({\cal H}) > 2$,
then a global valuation over $L({\mathcal H})$ is not possible.
\qed
\\
\end{theo}

It is also possible to give a topological version of this theorem
in the frame of local sections of sheaves. In fact, let $L$ be an
orthomodular lattice. We consider the family ${\cal W}_L$ of all
Boolean subalgebras of $L$ ordered by inclusion and the
topological space $\langle {\cal W}_L, {\cal W}_L^+ \rangle$. On
the set $$E_L = \{(W,f): W\in {\cal W},\ f:W \rightarrow  {\bf 2}\
\mbox{{\it f is a Boolean homomorphism}}  \} $$

\noindent we define a partial ordering given by $(W_1,f_1) \leq
(W_2,f_2)$ iff $W_1 \subseteq W_2$ and $f_1 = f_2 \mid W_1$. Thus
we can consider the topological space $\langle E_L, E_L^+ \rangle$
whose canonical base is given by the principal decreasing sets
$((W,f)] = \{(G ,f\mid G): G \subseteq W \}$. By simplicity
$((W,f)]$ is noted as $(W,f]$.

\begin{definition}
{\rm The map $p_L:E_L \rightarrow {\cal W}_L$ such that $(W,f)
\mapsto W$ is a sheaf over ${\cal W}_L$ called {\it spectral
sheaf} associated to the orthomodular lattice $L$. }
\end{definition}

Let $\nu: U \rightarrow E_L$ be a local section of $p_L$. By
(Domenech and Freytes, 2005; Proposition 4.2), for each $W\in U$
we have that $\nu(W) = (W,f)$ for some Boolean homomorphism $f:W
\rightarrow {\bf 2}$ and if $W_0 \subseteq W$, then $\nu(W_0) =
(W_0, f\mid W_0)$. From a physical perspective, we may say that
the spectral sheaf takes into account the whole set of possible
ways of assigning truth values to the propositions associated with
the projectors of the spectral decomposition ${\bf A} = \sum_{i}
a_{i} {\bf P}_i$. The continuity of a local section of $p$
guarantees that the truth value of a proposition is maintained
when considering the inclusion of subalgebras. In this way, the
compatibility condition  \ref{ccon} of the Boolean valuation with
respect to the intersection of pairs of Boolean sublattices of
$L({\mathcal H})$ is maintained. Thus, continuous local sections
of $p_L$ are identifiable to compatible contextual valuations.

We use $\nu(a) = 1$ to note that there exists $W\in U$ such that
$a\in W$, $\nu(W) = (W,f]$ and $f(a) = 1$. On the other hand, if
$f:W\rightarrow 2$ is a Boolean homomorphism, $\nu:(W]\rightarrow
E_L$ is such that for each $W_i \in (W]$, $\nu(W_i) = (W_i,
f/W_i)$ is a local section of $p_L$. We call this a  {\it
principal} local section.

A global section $\tau: {\cal W}_L \rightarrow E_L $ of $p_L$ is
interpreted as follows: the map assigns to every $W \in {\cal
W}_L$ a fixed Boolean valuation $\tau_w:W \rightarrow {\bf 2}$
obviously satisfying the compatibility condition. Thus, KS theorem
in terms of the spectral sheaf reads (Domenech and Freytes, 2005;
Theorem 4.3):

\begin{theo}\label{CS}
If $\mathcal{H}$ is a Hilbert space such that $dim({\cal H}) > 2$
then the spectral sheaf $p_{L(\mathcal{H})}$ has no global sections.
\qed
\end{theo}

\section{An algebraic study of modality}

With these tools, we are now able to build up a framework to
include modal propositions in the same structure as actual ones.
To do so we enrich the orthomodular lattice with a modal operator
taking into account the following considerations: 1) Propositions
about the properties of the physical system are interpreted in the
orthomodular lattice of closed subspaces of the Hilbert space of
the (pure) states of the system.  \hspace{0.2cm} 2) Given a
proposition about the system, it is possible to define a context
from which one can predicate with certainty about it together with
a set of propositions that are compatible with it and, at the same
time,  predicate probabilities about the other ones. In other
words, one may predicate truth or falsity of all possibilities at
the same time, i.e. possibilities allow an interpretation in a
Boolean algebra. In rigorous terms, for each proposition $P$, if
we refer with $\Diamond P$ to the possibility of $P$, then
$\Diamond P$ will be a central element of the orthomodular
structure. \hspace{0.2cm} 3) If $P$ is a proposition about the
system and $P$ occurs, then it is trivially possible that $P$
occurs. This is expressed as $P \leq \Diamond P$. \hspace{0.2cm}
4) Assuming an actual property and a complete set of properties
that are compatible with it determines a context in which the
classical discourse holds. Classical consequences that are
compatible with it, for example probability assignments to the
actuality of other propositions, shear the classical frame. These
consequences are the same ones as those which would be obtained by
considering the original actual property as a possible one. This
is interpreted in the following way: if $P$ is a property of the
system, $\Diamond P$ is the smallest central element greater than
$P$. From consideration 1) it follows that the original
orthomodular structure is maintained. The other considerations are
satisfied if we consider a modal operator $\Diamond$ over an
orthomodular lattice $L$  defined as $\Diamond a = Min \{z\in
Z(L): a\leq z \}$ with $Z(L)$ the center of $L$.

Let $A$ be an orthomodular lattice. We say that $A$ is {\it
Boolean saturated}  if and only if for all $a\in A$ the set
$\{z\in Z(A): z\leq a \}$ has a maximum (Domenech and Freytes,
2005). In this case, the maximum is denoted by $\Box (a)$. In view
of (Maeda and Maeda, 1970; Lemma 29.16), complete orthomodular
lattices with an operator $e(a) = \bigvee \{z \in Z(L) : z \leq a
\}$, are examples of Boolean saturated orthomodular lattices. They
form a variety of algebras $ \langle A, \land, \lor, \neg, \Box,
0, 1 \rangle$ of type $ \langle 2, 2, 1, 1, 0, 0 \rangle$, noted
as ${\cal OML}^\Box$ (Domenech {\it et al.}, 2006). ${\cal
OML}^\Box$ are axiomatized as follows:

\begin{enumerate}

\item[]
S1 \hspace{0.2cm}Axioms of ${\cal OML}$ \hspace{0.5cm} S5 \hspace{0.2cm} $\Box(x \land y) = \Box(x) \land \Box(y)$

\item[]
S2 \hspace{0.2cm}$\Box x \leq x$ \hspace{2.15cm}  S6 \hspace{0.2cm}  $y = (y\land \Box x) \lor (y \land \neg \Box x)$

\item[]
S3 \hspace{0.2cm}$\Box 1 = 1$ \hspace{2.2cm} S7 \hspace{0.2cm} $\Box (x \lor \Box y ) = \Box x \lor \Box y $

\item[]
S4 \hspace{0.2cm}$\Box \Box x = \Box x$ \hspace{1.6cm} S8 \hspace{0.2cm} $\Box (\neg x \lor (y \land x)) \leq \neg \Box x \lor \Box y $

\end{enumerate}

On each algebra of ${\cal OML}^\Box$ we can define the {\it
possibility operator} as the unary operation $\Diamond$ given by
$\Diamond x = \neg \Box \neg x$. It satisfies $a\leq \Diamond a$
and $\Diamond a = Min \{z\in Z(A): a\leq z  \}$. If $L$ is an
orthomodular lattice then there exists an orthomodular
monomorphism $f:L \rightarrow A$ such that $A \in {\cal OML}^\Box$
(Domenech {\it et al.}, 2006; Theorem 10). We refer to $A$ as a
{\it modal extension} of $L$. In this case we can see the lattice
$L$ as a subset of $A$. If $L^\Box \in {\cal OML}^\Box$ is a modal
extension of $L$, we define the {\it possibility space} of $L$ in
$L^\Box$ as $\Diamond L  = \langle \{\Diamond p : p \in L \}
\rangle_{L^\Box}$. If $W$ is a Boolean sublattice of $L$ then
$\langle W \cup \Diamond L \rangle_{L^\Box}$ is a Boolean
sublattice of $L^\Box$; in particular $\Diamond L$ is a Boolean
sublattice of $Z(L^\Box)$ (Domenech {\it et al.}, 2006; Theorem
14). The possibility space represents the modal content added to
the discourse about properties of the system.

\section{Sheaves and modality}
Let us  consider $L^\Box$ a modal extension of $L$. Then, the
spectral sheaf $p_L$ is a subsheaf of $p_{L^\Box}$. In this case
we refer to $p_{L^\Box}$ as a {\it modal extension} of $p_L$. It
is clear that local sections of $p_L$ can be seen as local
sections of $p_{L^\Box}$. We define the set $$Sec (\Diamond L) =
\{\nu:(\Diamond L] \rightarrow E_{L^\Box}: \nu \hspace{0.2cm} is
\hspace{0.2cm} principal \hspace{0.2cm} section \hspace{0.2cm} of
\hspace{0.2cm} p_{L^\Box} \}$$

Since $\Diamond L$ is a Boolean algebra, it is a subdirect product
of $\bf 2$. Thus, it always exists a Boolean homomorphism
$f:\Diamond L \rightarrow 2$, resulting $Sec (\Diamond L) \not=
\emptyset$. From a physical point of view, $Sec (\Diamond L)$
represents all physical properties as  possible properties. The
fact that $Sec (\Diamond L) \not= \emptyset$ shows that, {\it in
the frame of possibility}, one may talk simultaneously about all
physical properties.

In the orthomodular lattice of the properties of the system, it is
always possible to choose a context in which any possible property
pertaining to this context can be considered as an actual one. We
formalize this fact in the following definition and then we prove
that this is always possible in our modal structure:

\begin{definition}
Let $L$ be an orthomodular lattice, $W$ a Boolean sublattice of
$L$, $q\in W$ and ${L^\Box}$ be a modal extension of $L$. If $\nu
\in Sec (\Diamond L) $ such that $\nu(\Diamond q) = 1$ then an
actualization of $q$ compatible with $\nu$ is an extension $\nu':U
\rightarrow E_{L^\Box}$ such that $(\langle W \cup \Diamond L
\rangle_{L^\Box}] \in U$
\end{definition}

\begin{theo}\label{ACT}
Let $L$ be an orthomodular lattice, $W$ a Boolean sublattice of
$L$, $q\in W$ and ${L^\Box}$ be a modal extension of $L$. If $\nu
\in Sec (\Diamond L) $ such that $\nu(\Diamond q) = 1$ then there
exists an actualization of $q$ compatible with $\nu$.
\end{theo}

\begin{proof}
Suppose that $\nu(\Diamond L) = (\Diamond L, f)$. Let $F$ be the
filter associated with the Boolean homomorphism $f$. We consider
the $\langle W \cup \Diamond L \rangle_{L^\Box}$-filter $F_q$
generated by $F\cup\{q\}$. $F_q$ is a proper filter. In fact: if
$F_q$ is not proper, then there exists $a \in F$ such that $a\land
q \leq 0$. Thus $q\leq \neg a$ being $\neg a$ a central element.
But $\Diamond q$ is the smallest Boolean element greater than $q$.
Then $\Diamond q \leq \neg a$ or equivalently $\Diamond q \land a
= 0$. And this is a contradiction since $\Diamond q,\ a \in F$.
Thus, we may extend $F_q$ to be a maximal filter $F_M$ in $\langle
W \cup \Diamond L \rangle_{L^\Box}$. If we consider the natural
projection $f_{F_M}: \langle W \cup \Diamond L \rangle_{L^\Box}
\rightarrow \langle W \cup \Diamond L \rangle_{L^\Box}/F_M \approx
{\bf 2}$, the local section $\nu': (\langle W \cup \Diamond L
\rangle_{L^\Box}]\rightarrow E_{L^\Box}$ is an actualization of
$q$ compatible with $\nu$. \qed
\end{proof}

The next theorem allows a representation of the Born rule in terms
of continuous local sections of sheaves. This rule quantifies
possibilities from a chosen spectral algebra.

\begin{theo}
Let $L$ be an orthomodular lattice, $W$ a Boolean sublattice of
$L$, and $\nu:(W]\rightarrow E_L $ a principal local section. If
we consider a modal extension ${L^\Box}$ of $L$ then there exists
an extension $\nu':U \rightarrow E_{L^\Box}$ such that $\langle W
\cup \Diamond L \rangle_{L^\Box} \in U$.
\end{theo}

\begin{proof}
Suppose that $\nu(W) = (W,f)$. Let $i: W \rightarrow \langle W
\cup \Diamond L \rangle_{L^\Box}$ be the Boolean canonical
embedding. We see that there exists a Boolean homomorphism $f':
\langle W \cup \Diamond L \rangle_{L^\Box} \rightarrow {\bf 2}$
such that $f = f'i = f'\mid W$ since ${\bf 2}$ is injective in the
variety of Boolean algebras (Sikorski, 1948). Thus $\nu': (\langle
W \cup \Diamond L \rangle_{L^\Box}] \rightarrow E_{L^\Box}$ is the
extension required. \qed
\end{proof}

We note that this reading of the Born rule is a kind of the
converse of the possibility of actualizing properties given by
Theorem \ref{ACT}.

\begin{definition}
{\rm Let $L$ an orthomodular lattice, $L^\Box$ be a modal
extension and $\nu \in Sec (\Diamond L)$. An actualization
compatible with $\nu$ is a global section $\tau: {\cal W}_L
\rightarrow E_L$ } of $p_L$ such that $\tau(W \cap \Diamond L) =
\nu(W \cap \Diamond L)$.
\end{definition}

\begin{theo}
Let $L$ be an orthomodular lattice. Then $p_L$ is a global section
$\tau$ iff for each modal extension ${L^\Box}$ there exists  $\nu
\in Sec (\Diamond L)$ such that $\tau$ is a compatible
actualization of $\nu$.
\end{theo}

\begin{proof}
Suppose that $p_L$ admits a global section $\tau: {\cal W}_L
\rightarrow E_L$ and let $\tau(W) = (W, f_W)$. We consider the
family $(A_W = W \cap \Box L)_{W \in {\cal W}_L}$. Let $f_0:
\bigcup_W A_W \rightarrow {\bf 2}$ such that $f_0(x) = f_W(x)$ if
$x\in W$. $f_0$ is well defined since $\tau$ is a global section.
If we consider $\langle \bigcup_W A_W \rangle_{L^\Box}$ the
subalgebra of $L^\Box$ generated by the join of the family
$(A_W)_W$, it may be proved that it is a Boolean subalgebra of
$\Diamond L$. We can extend $f_0$ to a Boolean homomorphism $f_0':
\langle \bigcup_W A_W \rangle_{L^\Box} \rightarrow {\bf 2}$.
 Since
${\bf 2}$ is injective in the variety of Boolean algebras
(Sikorski, 1948), then there exists a Boolean homomorphism $f:
\Diamond L \rightarrow {\bf 2}$ which extends it to  $f_0'$. If we
consider $\nu \in Sec (\Diamond L)$ such that $\nu(G) = (G, f\mid
G)$, it result that $\tau$ is a compatible actualization of $\nu$.
The converse is immediate. \qed
\end{proof}
\\

To conclude we may say that the addition of modalities to the
discourse about the properties of a quantum system enlarges its
expressive power. At first sight it may be thought that this could
help to circumvent contextuality, allowing to refer to physical
properties belonging to the system in an objective way that
resembles the classical picture. In view of the last theorem,
since any global section of the spectral sheaf is a compatible
actualization of a local one belonging to $Sec (\Diamond L)$, a
global actualization that would correspond to a family of
compatible valuations is prohibited. Thus, the theorem states that
the contextual character of quantum mechanics is maintained even
when the discourse is enriched with modalities.\\

\noindent {\bf Acknowledgements} \noindent This work was partially
supported by the following grants: PICT 04-17687 (ANPCyT), PIP
N$^o$ 6461/05 (CONICET), UBACyT N$^o$ X081 and X204.

\end{document}